# Stability and Carrier Transport Properties of Phosphorene Based Polymorphic Nanoribbons

Sumandeep Kaur[1,2], Ashok Kumar[2*], Sunita Srivastava[1], Ravindra Pandey[3] and K. Tankeshwar[1,4*]

[1]Department of Physics, Panjab University, Chandigarh 160014, India

[2]Department of Physical Sciences, School of Basic and Applied Sciences, Central University of Punjab, Bathinda, 151001, India

[3]Department of Physics, Michigan Technological University, Houghton, MI, 49931, USA

[4]Department of Physics, Guru Jambheshwar University of Science and Technology, Hisar, 125001, Haryana, India

(January 27, 2018)

*Emails:

Ashok Kumar (ashokphy@cup.edu.in)

K. Tankeshwar (tkumar@gjust.org)

# Abstract

A few-layer black phosphorene has recently gained significant interest in the scientific community. In this paper, we consider several polymorphs of phosphorene nanoribbons (PNRs) and employ deformation potential theory within the effective mass approximation together with density functional theory to investigate their structural, mechanical and electronic properties. The results show that stability of PNRs strongly depends on the direction along which they can be cut from 2D counterpart. PNRs also exhibit a wide range of line stiffness ranging from $6 \times 10^{10}$ eV/m to $18 \times 10^{11}$ eV/m which has little dependence on the edge passivation. Likewise, the calculated electronic properties of PNRs display them to be either narrow-gap semiconductor ($E_g < 1$ eV) or wide-gap semiconductor ($E_g > 1$ eV). The carrier mobility of PNRs is found to be comparable to that of the black phosphorene. Some of the PNRs show n-type (p-type) semiconducting character owing to their higher electron (hole) mobility. Passivation of the edges leads to n-type ↔ p-type transition in many of the PNRs considered. The predicted novel characteristics of PNRs with a wide range of mechanical and electronic properties make PNRs to be potentially suitable for the use in nanoscale devices.

# 1. Introduction

Since its successful exfoliation in 2014, a few-layer black phosphorene has gained significant interest in the scientific community [1]. Monolayer black phosphorene (α-P) with its unique anisotropic puckered structure is a direct band gap semiconductor with a bandgap of ~ 1.0 eV [2]. It is found to withstand mechanical strains as high as 40% with a large lateral flexibility [3]. Another equally stable two-dimensional (2D) allotrope of phosphorus is blue phosphorene (β-P) which is recently realized experimentally [4, 5]. Blue phosphorene possess graphene-like structure with the out-of-plane buckling of 2.2 Å and an indirect band gap of ~2.0 eV [6]. Both black and blue phosphorene have intrinsic carrier mobility as high as $10^3$ $cm^2V^{-1}s^{-1}$ [7, 8]. The high carrier mobility and semiconducting character of α- and β-P makes them promising material for the potential application in the optical and electronic devices [9].

Bulk phosphorus can form various allotropes such as violet, red, white and black due to the inequivalent $sp^3$ hybridization of orbitals in the lattice [3]. Likewise, several structurally different 2D polymorphs of phosphorous namely, α-P, β-P, γ-P, δ-P, ε-P, τ-P, η-P, θ-P, ϕ-P, tricycle-type red phosphorene (R-P), square-octagon phosphorene (O-P) and hexagonal-star phosphorene (H-P) have been investigated using first principles method [10-17]. α-P, β-P, γ-P, δ-P and R-P exhibit buckled honeycomb structure consisting of a six membered ring similar to graphene, while ε-P, τ-P, η-P, θ-P, ϕ-P, O-P, H-P crystallize into non-honeycomb structural arrangements. Red phosphorene is constructed by the in-plane connections of the segments of α-P and β-P [14]. ε-P and τ-P consist of squared units of phosphorus atoms while η-P and θ-P have phosphorus atoms in the pentagon structural arrangement [12]. ϕ-P consists of 4, 6 and 10 membered rings [13], and O-P allotrope contains a unique atomic octagonal tiling (OT) pattern consisting of 4 and 8 membered rings [15]. The phosphorus atoms in H-P forms a hexagonal lattice with a Magen-David-like top view [16]. The honeycomb structures of phosphorene exhibit two type of edges i.e., armchair and zigzag, whereas most of the non-honeycomb structures are found to possess more than two distinctly different edges. All these 2D polymorphs are found to be semiconducting in nature with band gap ranging from 0.4 eV to 2.1 eV [10-17]. Note that α-P and β-P are experimentally realized, whereas the other allotropes are yet to be synthesized.

Similar to one-dimensional (1D) graphene nanoribbons (GNRs), the phosphorene 1D nanoribbons can be constructed from their 2D counterpart by cutting along various edge direction. Several 1D structures of black and blue phosphorene have been investigated previously [9]. The α- and β-phosphorene nanoribbons possess two types of edges i.e. armchair (AC) and zigzag (ZZ), which strongly influence their band gaps [18, 19]. External electric field has been shown to modify electronic band gap of both AC and ZZ α-PNRs due to giant stark-effect [20, 21]. Also, the electron/hole effective masses and carrier mobility of both α- and β-PNRs are found to depend on the edge configurations [22]. On the other hand, in the non-honeycomb PNR structures, various types of edges exist and it will be interesting to investigate their properties.

In the present work, we consider 5 honeycomb and 8 non-honeycomb phosphorene polymorphs, and report the results of density functional theory calculations to determine energetics, mechanical stability and carrier mobility. In the following, Sec 2 briefly describes the computational model. Results are discussed in Sec 3, and summary is given in Sec 4.

## 2. Computational Method

Density functional theory calculations were performed using the SIESTA program package [23]. The norm-conserving Troullier Martin pseudopotential was used to treat the electron-ion interactions [24], whereas the exchange and correlation energies were described using the GGA functional form [25]. Double zeta basis set with polarization functions (DZP) were used to expand the Kohn Sham orbitals with the mesh cuttoff energy of 450 Ry. The minimization of energy was carried out using the conjugate-gradient (CG) technique with the forces less than 0.01 eV/Å on each atoms. Monkhorst-Pack scheme was used to sample Brillouin zone with a (30×30×1) mesh for 2D sheets. A (30×1×1) and (1×30×1) mesh was used for PNRs with length along $x$ and $y$ direction, respectively. A vacuum region greater than 20 Å along the two direction has been used in calculations to avoid the superficial interactions due to replicas.

The carrier transport in PNRs was calculated by the deformation potential (DP) theory [26] and effective mass approximation. The lowest energy PNR structures were considered for calculations. For one-dimensional (1D) systems, an analytical expression [27] for the mobility ($\mu$) was employed as follows:

$$\mu_{1D} = \frac{e\hbar^2 C_{1D}}{(2\pi k_B T)^{1/2} |m^*|^{3/2} E_1^2}, \qquad (1)$$

where $T = 300$ K and $m^*$ is the effective mass of the charge carriers, defined as $m^* = \hbar^2(\partial^2 E(k)/\partial k^2)^{-1}$. $C$ is the stretching modulus caused by the uniaxial-strain ($\varepsilon$), which has been calculated using the expression $C_{1D} = \frac{1}{L_0}\frac{d^2 E_S}{d\varepsilon^2}$, in which $E_S$ is the strain energy of a unit cell and $L_0$ is the equilibrium lattice constant. $E_1$ in Eqn. (1) is the deformation-potential (DP) constant, that denote the shift of the band-edge energy induced by strain, and has been obtained as $E_1 = \frac{dE_{edge}}{d\varepsilon}$, where $E_{edge}$ is the energy of the conduction band minimum (for electrons) or the valence band maximum (for holes). Note that Eqn. (1) has been successfully applied previously to study the intrinsic mobility of GNRs [28], graphyne nanoribbons [29] and PNRs [9].

## 3. Results and Discussions

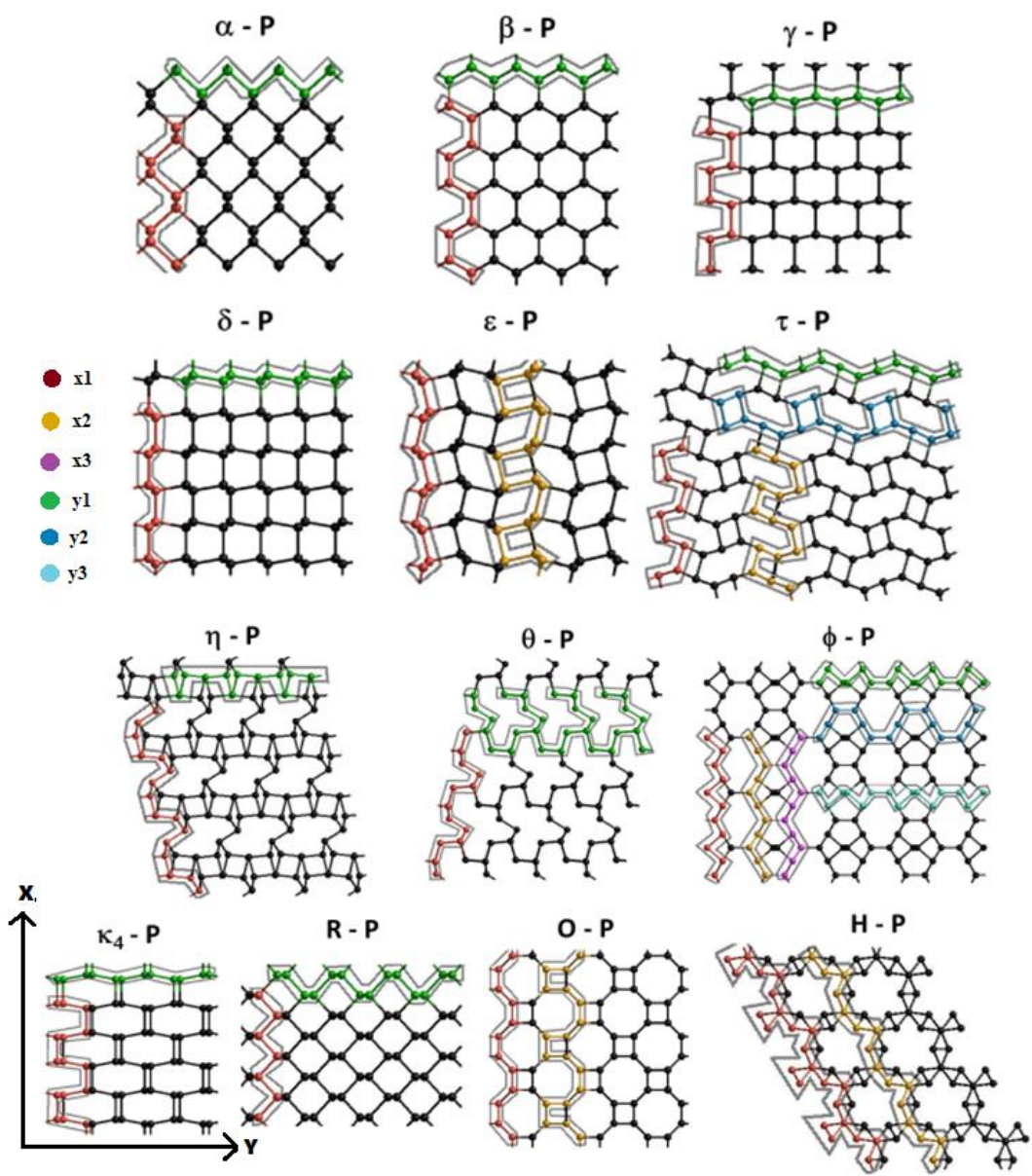

**FIGURE 1:** Phosphorene allotropes indicating different type of edge (with different colors) in different direction (*x* and *y*). Since η-P and θ-P possess large number of edge structures, a few representative cases are presented here, others are given in in Figure S1 of ESI.

Depending upon the ways to connect tetrahedrally coordinated P-atoms in a 2D lattice, different structural allotropes can be formed. We have considered several allotropes of phosphorene, including β-P, γ-P, δ-P, ε-P, τ-P, η-P, θ-P, φ-P, $κ_4$-P, R-P, O-P and H-P. All these allotropes are found to be semiconducting with calculated band gap varying from 0.48 eV to 2.09 eV (Table S1 of ESI).

The magnitude of the relative formation energies ($E_{RFmono}$) of all these 2D allotropes are found to be of the order of 10-10$^2$ meV/atom, indicating their ease of formation. This suggests that not only the α- and β-P, but the other allotropes can also be experimentally realized. Hence, it is worth to explore the properties of other allotropes. Note that $E_{RFmono}$ can be obtained as: $E_{RFmono} = \frac{E_A - NE_\alpha}{N}$, where $E_A$ is the total energy of the considered phosphorene allotrope and $E_\alpha$ is the energy per atom of the most stable phosphorene allotrope i.e. α-P and $N$ is the number of atoms per unit cell of the allotrope. Similarly, the cohesive energies are in the range of -5.50 to -5.66 eV/atom and are comparable to that of α-P (-5.68 eV/atom) (Table S1 of ESI).

Figure 1 shows the different types of possible edges of allotropes of phosphorene considered. Each allotrope cut along the distinct edges gives rise to a distinct phosphorene nanoribbon. The honeycomb structures (α-P, β-P, γ-P, δ-P and R-P) possess two edge structures one along x-direction and the other along y-direction. Unlike these honeycomb structures which possess only one type of edge along each direction, the non-honeycomb structures possess more than one different types of edges along different directions. Due to perfect square symmetry in the unit cell of ε-P and O-P, they possess two different edge structures ($\varepsilon_{x1}$, $\varepsilon_{x2}$ and $O_{x1}$, $O_{x2}$) along one direction which are also similar in other direction (Figure 1). The H-P allotrope also exhibit two types of edge structures ($H_{x1}$ and $H_{x2}$) along one direction which are also similar in other direction due to the hexagonal symmetry in its structure. τ-P forms four PNR structures corresponding to the two different type of edges along each direction shown in Figure 1. η-P possesses 11 PNR structures, 5 along x axis ($\eta_{x1}$, $\eta_{x2}$, $\eta_{x3}$, $\eta_{x4}$ and $\eta_{x5}$) and 6 along y axis ($\eta_{y1}$, $\eta_{y2}$, $\eta_{y3}$, $\eta_{y4}$, $\eta_{y5}$ and $\eta_{y6}$) (Figure 1 & Figure S1 of ESI). θ-P possesses 7 PNRs structures, 4 along x axis ($\theta_{x1}$, $\theta_{x2}$, $\theta_{x3}$ and $\theta_{x4}$) and 3 along y axis ($\theta_{y1}$, $\theta_{y2}$ and $\theta_{y3}$). ϕ-P form 6 PNR structures, 3 along x axis ($\phi_{x1}$, $\phi_{x2}$, and $\phi_{x3}$) and 3 along y axis ($\phi_{y1}$, $\phi_{y2}$ and $\phi_{y3}$). κ$_4$-P possesses two different structures one along each direction ($\kappa_{4x1}$ and $\kappa_{4y1}$). The width of considered PNRs lies in the range 17 Å to 46 Å [Table 1].

Therefore, depending on the edge configuration in x- and y-direction, total 46 structures [Figures 1, S1 and S2 of ESI] were considered. Note that, in the equilibrium configurations, the edge atoms in the bare PNRs show reconstruction due to the presence of the dangling bonds [Figure S2 of ESI]. The dangling bonds on the edges of a given PNR are then terminated by hydrogen atoms [Figure S3].

**TABLE 1.** Calculated deformation potential ($E_1$), energy bandgap of bare and passivated PNRs. (a) is length and (w) is width along the periodic direction of the bare PNRs.

| System | Length (a) (Å) | width (w) (Å) | Bare PNR | | | Passivated PNR | | |
|---|---|---|---|---|---|---|---|---|
| | | | $E_1$ (electrons) (eV) | $E_1$ (holes) (eV) | $E_g$ (eV) | $E_1$ (electrons) (eV) | $E_1$ (holes) (eV) | $E_g$ (eV) |
| $\alpha_{x1}$ | 4.46 | 18.99 | 1.35 | -1.20 | 0.37 | 1.61 | -1.5 | 0.99 |
| $\beta_{x1}$ | 5.81 | 17.39 | 4.51 | -4.10 | 1.08 | -7.05 | 0.14 | 2.20 |
| $\gamma_{x1}$ | 5.40 | 18.37 | 3.25 | -3.25 | 0.36 | 1.86 | -1.75 | 0.71 |
| $\delta_{x1}$ | 5.39 | 29.96 | 3.54 | -3.17 | 0.35 | 4.34 | -4.33 | 0.41 |
| $\varepsilon_{x1}$ | 5.40 | 29.09 | -0.70 | 0.14 | 0.64 | 0.76 | -0.83 | 0.57 |
| $\tau_{x1}$ | 5.33 | 38.10 | -4.97 | 4.47 | 0.73 | -6.74 | 2.14 | 1.27 |
| $\tau_{y2}$ | 6.53 | 29.64 | -1.08 | 2.27 | 1.14 | -2.21 | 0.91 | 1.27 |
| $\eta_{x3}$ | 6.39 | 25.39 | 0.80 | -0.66 | 0.74 | 2.03 | -2.04 | 0.99 |
| $\eta_{y2}$ | 5.43 | 33.73 | -1.66 | 1.52 | 0.85 | -3.92 | 2.41 | 0.99 |
| $\theta_{x1}$ | 6.36 | 31.12 | 0.46 | -0.68 | 0.16 | 1.75 | -1.74 | 0.75 |
| $\theta_{y3}$ | 5.54 | 30.36 | 0.05 | -0.70 | 1.08 | -0.58 | 1.58 | 1.25 |
| $\phi_{x1}$ | 6.20 | 46.35 | -4.03 | 3.05 | 0.32 | -0.81 | 0.91 | 1.03 |
| $\phi_{y3}$ | 7.81 | 37.53 | 0.49 | -0.56 | 0.90 | -0.91 | 1.22 | 1.04 |
| $\kappa_{4x1}$ | 5.52 | 30.40 | 4.64 | -2.06 | 0.30 | 1.81 | -1.92 | 0.39 |
| $R_{y1}$ | 8.92 | 18.96 | 2.45 | -2.40 | 0.51 | -3.12 | -9.21 | 1.26 |
| $O_{x2}$ | 6.53 | 34.71 | -4.71 | -4.96 | 1.81 | -2.43 | -0.62 | 2.10 |
| $H_{x1}$ | 6.27 | 37.09 | 2.28 | 3.81 | 1.40 | 1.83 | 3.44 | 1.48 |

## 3.1 Structural Stability

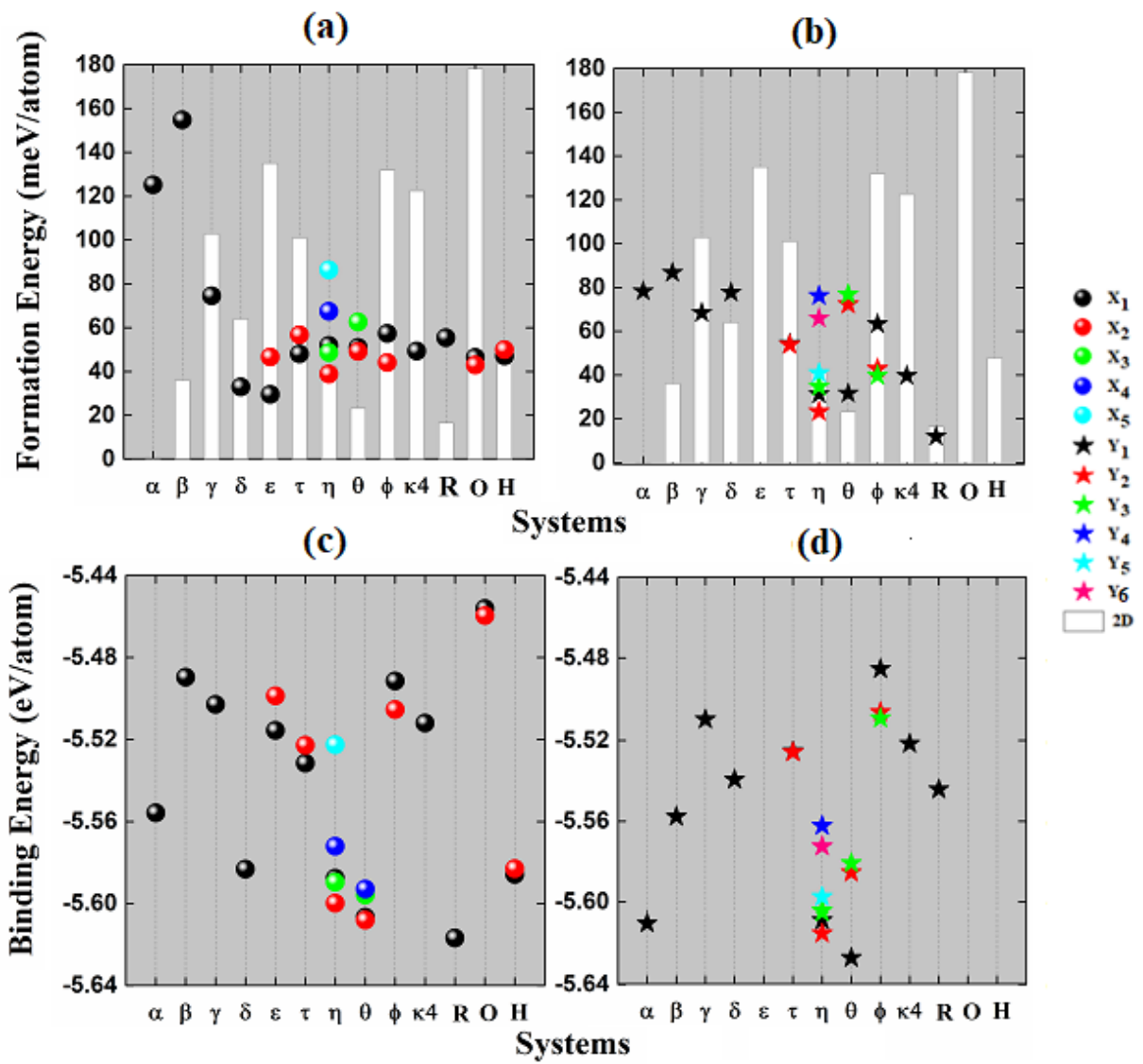

**FIGURE 2:** Formation energy per atom and binding energy per atom of the PNRs considered. The bars in (a) and (b) shows the data for the 2D sheet.

In order to study the energetic stability of the considered PNRs, we have calculated their formation energy (E$_{FPNR}$) which is defined as the energy required to form these nanoribbons from their 2D counterpart. The formula is given as: $E_{FPNR} = \frac{E_R - NE_{2D}}{N}$, where $E_R$ is the total energy of the ribbons, $E_{2D}$ is the energy per atom of the 2D sheet and $N$ is the number of atoms per unit cell of the ribbons. $E_{FPNR}$ of ribbons (except β-PNRs) are found to be lower than that of the black phosphorene nanoribbons (α-PNRs), thereby indicating these PNRs to be energetically more favourable than the PNRs of most stable black phosphorus.

The energetic stability of PNRs are found to be strongly dependent on the direction along which they can be cut from their 2D counterpart. It is found that four of the 2D monolayer structures (i.e. β-P, η-P, θ-P and R-P), have formation energy comparable (the energy difference being less than 40 meV/atom) with the most stable monolayer structure (α-P) [Figures 2 (a & b)], indicating these monolayers to be equally stable. Other 2D structures are energetically not favourable, however, their 1D counterparts show energetically favourable structures which is evident from their low value of formation energies. Note that β-P, η-P, θ-P, R-P and H-P possesses least tension in their structures due to the presence of either 5 or 6 membered rings whereas the increased tension in ε-P, τ-P, ϕ-P, κ$_4$-P and O-P due to the presence of 4 membered rings in their structures makes them less favourable. However, the energy difference of the order of a few meV indicates their ease of formation.

In order to get further insight into energetics of given PNRs structures, the binding energy was calculated using the formula: $E_B = \frac{E_R - NE_S}{N}$, where $E_R$ is the total energy of the ribbon, $E_S$ is the energy of the isolated single phosphorene atom and $N$ is the number of atoms in ribbons. The binding energies of the PNRs lies within the range -5.65 to -5.45 eV/atom [Figures 2 (c & d)] which are comparable with the binding energy (-5.68 eV/atom) of most stable black phosphorene monolayer.

## 3.2 Mechanical Property

The representative structures in each edge direction based on the minimum total energy criterion were considered to determine their mechanical property which led to total of 34 semiconducting (17 bare and 17 passivated) PNRs. Stiffness of a material is an important parameter to describe its mechanical stability. The stiffer an object is, the less flexible it is. In 1D ribbons one can calculate line stiffness using the formula: $C_{1D} = \frac{1}{L_0}\frac{d^2 E_S}{de^2}$, where $E_S$ is the difference in the total energy of the equilibrium and the strained PNRs, $e$ is the applied strain and $L_0$ is length of the ribbon. For each ribbon, a strain varying from -1% to +1% in the steps of 0.5% is applied along the length of the ribbon. The line stiffness is then calculated by fitting the strained energy ($E_S$) versus strain ($e$) curve with the formula $E_S = a_0 e^2 + a_1 e + a_2$ [Figure 3 (a & b) and Figure S4].

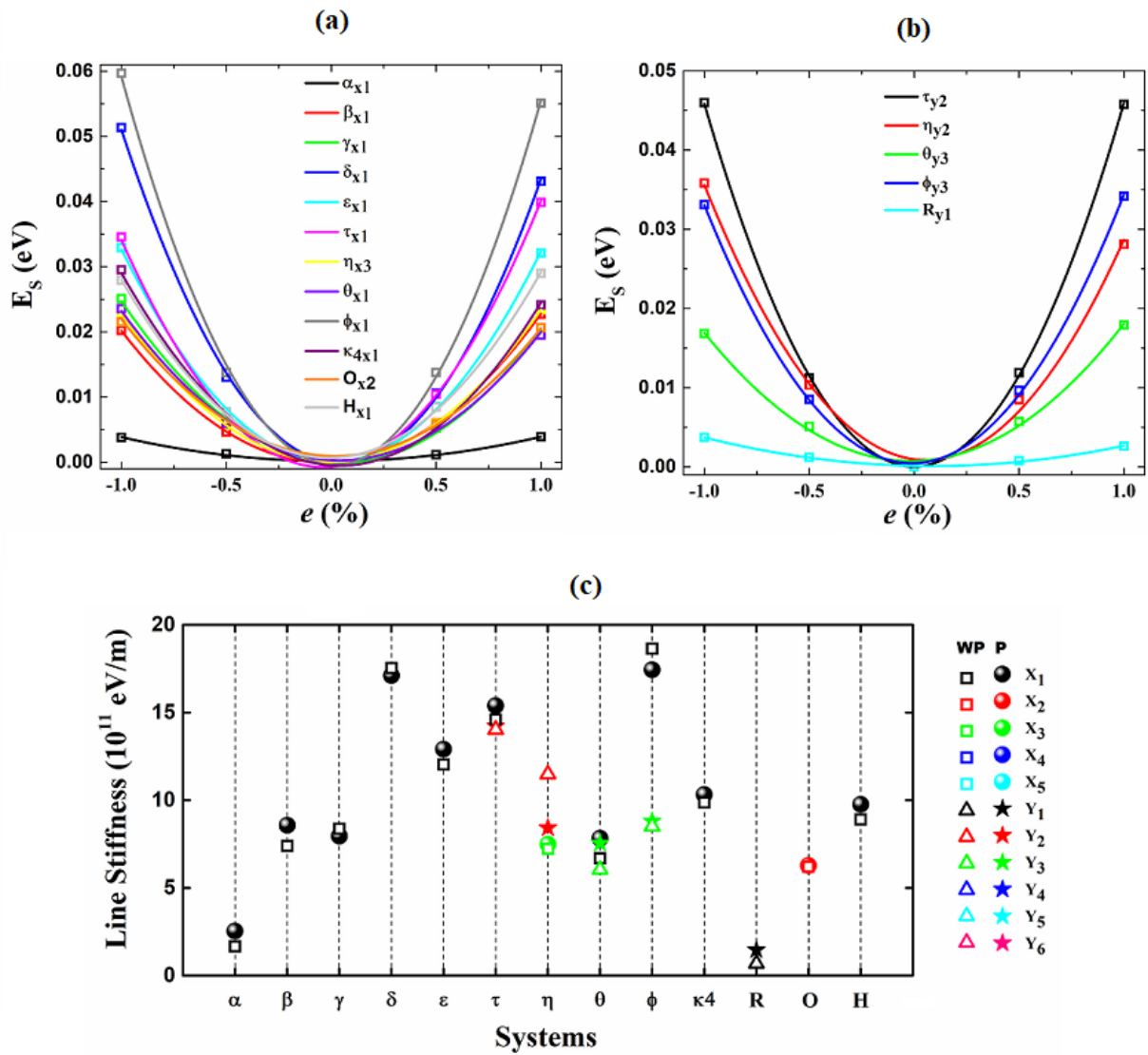

**FIGURE 3: (TOP)** Bare PNRs: strain energy vs. strain applied along (a) x-axis, and (b) y-axis. **(BOTTOM)** (c) The line stiffness of bare (WP) and passivated (P) PNRs considered.

It is found that the line stiffness of PNRs varies from $6 \times 10^{10}$ eV/m (for $R_{y1}$-PNR) to $18.6 \times 10^{11}$ eV/m (for $\phi_{x1}$-PNR) [Figure 3] indicating R-PNR cut along y-direction to be most flexible while ϕ-P cut along x direction to be the least flexible. Note that the calculated value of α-PNR is $16.6 \times 10^{10}$ eV/m. Line stiffness depends upon the inverse of the length of the nanoribbon, which does not get effected on passivation. Also, the strain is applied in the small region i.e. ±1% where hooks law is applicable. The rate of change of energy with strain for the passivated and unpassivated PNRs remain nearly identical that leads to nearly same line stiffness for both types of nanoribbons [Figures 3] and their mechanical flexibility does not change much with the passivation of edges.

### 3.3 Electronic and Carrier Transport Properties

Next, we calculate the electronic structure of the PNRs considered. The semiconducting behavior of given PNRs varies from narrow-gap semiconductor ($E_g < 1$ eV) to wide-gap semiconductor ($E_g > 1$ eV). Passivation of edges with hydrogen increases the band gap of PNRs which lies in the range 0.3-2.2 eV [Table 1 and Figures 4]. The increase in bandgap on the edge-passivation is attributed to the pairing of electrons due to hydrogen atoms which eliminates the dangling bonds. On passivation, direct↔indirect band gap transition has been found to occur in these PNRs. For example, in $\alpha_{x1}$-PNR, CBM shifts from a point between Γ&X to Γ while VBM remains at Γ; in $\gamma_{x1}$-PNR, CBM shifts from X to Γ while VBM remains at Γ; in $\phi_{x1}$-PNR, CBM remains at Γ while VBM shifts from X to Γ leading to an (indirect→direct transition) in these PNRs.

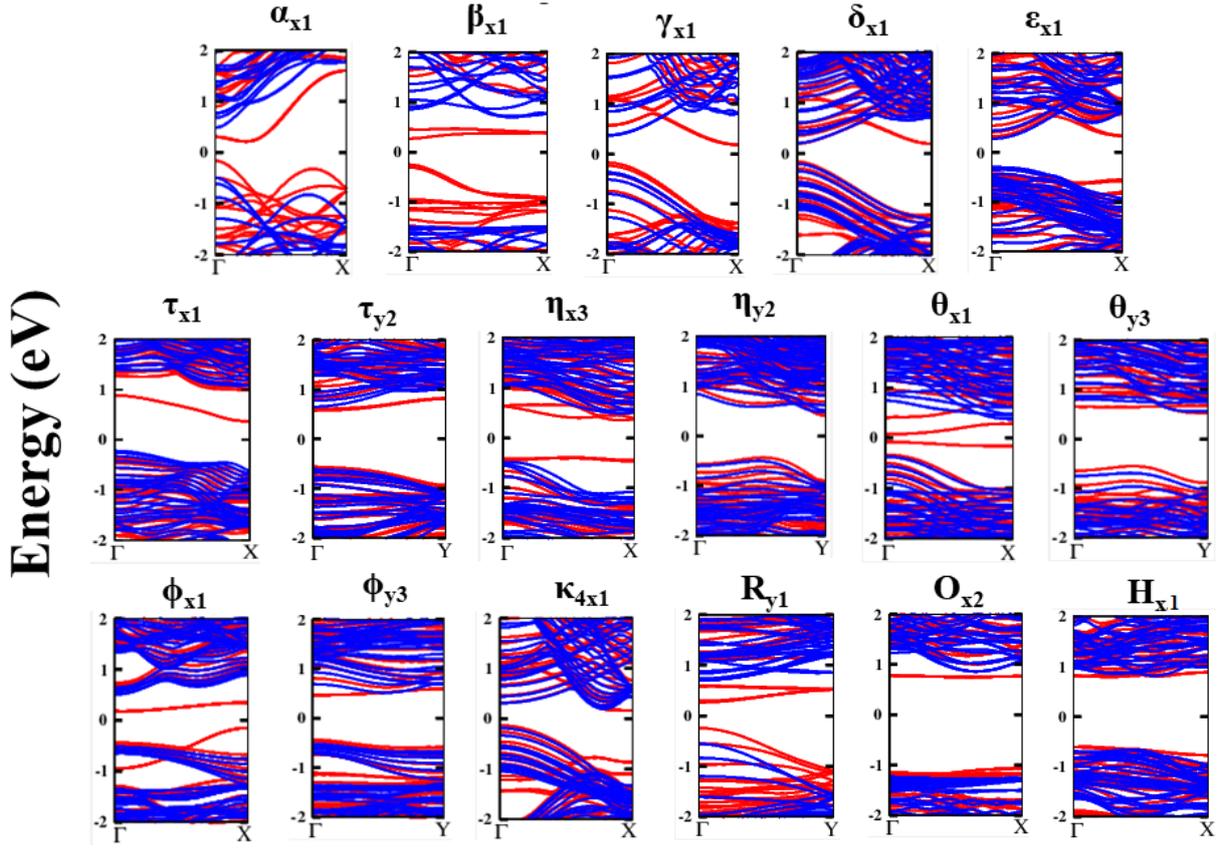

**FIGURE 4:** Calculated band structures of phosphorene nanoribbons. Red color indicates bands of the bare PNRs while passivated PNRs are shown by blue color.

On the other hand, direct→indirect bandgap transition occurs in $\beta_{x1}$-PNR due to a shift of CBM from Γ to a point between Γ&X and VBM from Γ to X, while in $\theta_{x1}$- PNR and $\phi_{y3}$-PNR this transition occurs only due to a shift of CBM from Γ to X and from Γ to a point between Γ&Y, respectively. The magnitude of band gap is found to be highly anisotropic, i.e., it is less for a PNR along the x direction compared to a PNR along y direction of same allotrope, for example, the bandgap of $\tau_{x1}$ ($E_g$ = 0.73 eV) and $\phi_{x1}$ ($E_g$ = 0.32 eV) is respectively less than the bandgap of $\tau_{y2}$ ($E_g$ = 1.14) and $\phi_{y3}$ ($E_g$ = 0.90). (Table 1).

To calculate the carrier mobility, the effective masses of electron or holes need to be determined [Eqn (1)]. The electron (hole) effective mass is inversely proportional to the curvature of the CBM (VBM) of a given band structure. Therefore, the effective mass of an electron (holes) can be calculated by fitting a small section of the E vs. *k* surface (Figure 4) in the vicinity of the CBM (VBM) at zero strain. The calculated results find a large value (2.26 $m_e$) of the electron effective mass in the bare $\kappa_{4x1}$-PNR which can be attributed to the presence of a flat conduction band (band in red colour at X-point in Figure 4) arising due to the dangling electron of the edge atoms. On passivation with hydrogen atoms, the band due to the dangling electrons disappear, and CBM possess larger curvature that gives rise to a small effective mass (0.04 $m_e$). The free charge carriers at the edges of unpassivated PNRs, gets bonded to hydrogen atoms, which effects their effective masses and hence their mobilities.

The calculated carrier effective masses are found to be anisotropic in nature [Figures 5]. Carrier effective masses of α-PNR ($m_e^*$ = 0.12 $m_e$ and $m_h^*$ = 0.08 $m_e$) and R-PNR ($m_e^*$ = 0.23 $m_e$ and $m_h^*$ = 0.08 $m_e$) are calculated to be lowest (close to zero line). The effective masses of most of the PNRs are comparable with α-PNR and R-PNR [Figures 5]. On passivation, the effective mass of most of the PNRs is found to be less than 0.3 $m_e$. Note that the carrier mobility is inversely proportional to 3/2 power of the effective mass, hence lower effective mass leads to an increase in the mobility.

Another factor which also determine the carrier mobility is the deformation potential (DP) [Eqn 1]. DP is calculated by the linear fitting of the valence band edge (VBE)/conduction band edge (CBE) vs. strain surface. The magnitude of DP describes the change in energy of the electronic band with the elastic deformation and, therefore, describes the degree to which the charge carriers interact with phonons. A lower value of DP indicates a weaker electron-phonon coupling in the conduction (valence) band, thereby contributing to an increase in the mobility of electrons (holes) [30]. The values of DP are listed in Table1. Note that the negative DP indicates the deformation in band edge to be in the opposite direction on the application of a strain.

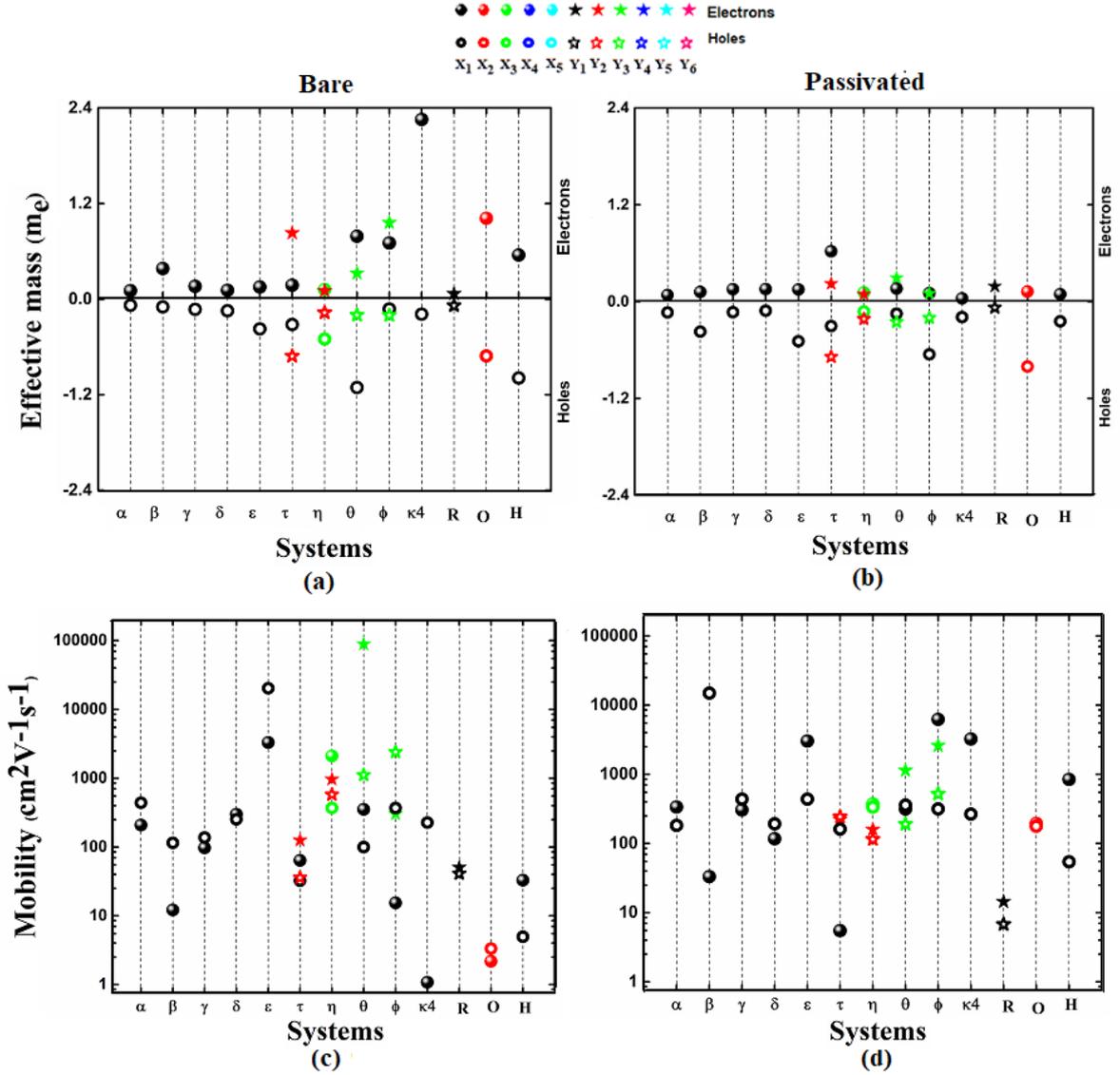

**FIGURE 5:** The effective masse and mobility of electrons and holes for bare and passivated PNRs at T= 300 K.

We now calculate the carrier mobility using Eqn 1. Our results find carrier's mobility of given PNRs to be highly anisotropic, e.g, $\mu_e$ ($\mu_h$) of $\theta_{y3}$ is 253 (11) times that of $\mu_e$ ($\mu_h$) of $\theta_{x1}$ and, $\mu_e$ ($\mu_h$) of $\phi_{y3}$ is 20 (7) time that of $\mu_e$ ($\mu_h$) of $\phi_{x1}$. Note that an anisotropy in the carrier mobility is also reported for black ($\alpha$) and blue ($\beta$) PNRs [7, 19, 31]. It is important to mention here that we have considered fixed width of PNRs, though the mobility also depends on the width of given ribbons [18, 32]. Although, line stiffness remains almost unaffected by the passivation of the edges but the effective masses and deformation potentials show significant modulation that leads to enhancement in the carrier mobility on edge-passivation, e.g., hole mobility in $\beta$-PNR increases 130 times and electron mobility in $\kappa_4$-PNR increases by 3000 times on passivation [Figure 5 and Table S2].

The carrier mobility of most of the PNRs considered are found to be comparable with the monolayer black phosphorene (i.e. $10^3$ cm$^2$V$^{-1}$s$^{-1}$) with the exception of $\beta$-PNR, $\tau$-PNR, R-PNR, and H-PNR. Particularly, the electron mobility of $\theta$-PNR and the hole mobility of $\varepsilon$-PNR is found to be one order higher than $\alpha$-PNR [Figure 5 and Table S1of ESI]. The electron mobility of the passivated PNRs including $\alpha_{x1}$, $\varepsilon_{x1}$, $\eta_{y2}$, $\theta_{y3}$, $\phi_{x1}$, $\phi_{y3}$, $\kappa_{4x1}$, $R_{y1}$, $O_{x2}$, $H_{x1}$ is higher than their hole mobility indicating their n-type semiconducting characteristic whereas $\beta_{x1}$, $\gamma_{x1}$, $\delta_{x1}$, $\tau_{x1}$, $\tau_{y2}$, $\eta_{x3}$, $\theta_{x1}$ behaves like p-type semiconductors owing to their higher hole mobility. Among the PNRs considered, $\alpha_{x1}$, $\delta_{x1}$, $\varepsilon_{x1}$, $\tau_{x1}$, $\tau_{y2}$, $\eta_{x3}$, $\theta_{x1}$, $\phi_{x1}$, $\phi_{y3}$, $\kappa_{4x1}$ and $O_{x2}$ PNR changes their carrier mobility, thereby, the semiconducting character (n-type ↔ p-type) after the passivation of the edges with H-atoms. Our results indicate that the intrinsic band gap and high carrier mobility possessed by these PNRs may find various applications in nano- and opto-electronics.

## 4. Summary

In summary, stability, electronic and mechanical properties of nanoribbons of 13 phosphorene allotropes namely α-P, β-P, γ-P, δ-P, ε-P, τ-P, η-P, θ-P, ϕ-P, κ$_4$-P, R-P, O-P, H-P have been investigated. The magnitude of formation energies of PNRs considered is found to be comparable to their monolayer counterparts indicating their ease of formation. Deformation potential theory within the effective mass approximation have been employed to the most stable semiconducting bare and passivated PNRs to analyse their carrier transport properties. On passivation of edges, band gap increases along with direct↔indirect transition in some cases. The effective mass of most of the PNRs is less than 0.3 m$_e$. The carrier mobility of most of the PNRs is found to be comparable to that of the monolayer α-P. The unique features of the PNRs considered render them as favourable 1D materials to be used in applications relating to phosphorene based devices.

## Acknowledgements

SK is grateful to UGC-BSR for financial assistance in the form of senior research fellowship. The computational facility at Central University of Punjab, Bathinda and RAMA High Performance Computing Cluster at Michigan Technological University Houghton, USA, are used to obtaining the results presented in this paper.